\documentclass[pdftex]{sigchi}
\pdfoutput=1

\usepackage{enumitem}
\usepackage{balance}  
\usepackage{graphics} 
\usepackage{times}    
\usepackage{url}      
\usepackage{subfigure}

\makeatletter
\def\url@leostyle{%
  \@ifundefined{selectfont}{\def\UrlFont{\sf}}{\def\UrlFont{\small\bf\ttfamily}}}
\makeatother
\def\pprw{8.5in}
\def\pprh{11in}

\usepackage{xcolor}

\newcommand{\Fig}[1]  {Figure~\ref{fig:#1}}

\newcommand{\Table}[1]  {Table~\ref{tbl:#1}}

\renewcommand \paragraph[1] {\vspace{0.05cm} \textbf{#1}}

\newcommand{\todo}[1]{\textcolor{magenta}{{[TODO: #1]}}}

\usepackage[pdftex]{hyperref}
\hypersetup{
    pdftitle={},
    pdfauthor={LaTeX},
    pdfkeywords={SIGCHI, proceedings, archival format},
    bookmarksnumbered,
    pdfstartview={FitH},
    colorlinks,
    citecolor=black,
    filecolor=black,
    linkcolor=black,
    urlcolor=black,
    breaklinks=true,
}

\begin{document}

\title{CamSwarm: Instantaneous Smartphone Camera Arrays \\ for Collaborative Photography}

\numberofauthors{3}
\author{
  \alignauthor Yan Wang \\
    \affaddr{Columbia University}\\
    \affaddr{2960 Broadway, New York, NY, USA}\\
    \email{yanwang@ee.columbia.edu}\\
  \alignauthor Jue Wang \\
    \affaddr{Adobe Systems}\\
    \affaddr{801 N 34th Street, Seattle, WA, USA}\\
    \email{juewang@adobe.com}\\
  \alignauthor Shih-Fu Chang\\
    \affaddr{Columbia University}\\
    \affaddr{2960 Broadway, New York, NY, USA}\\
    \email{sfchang@ee.columbia.edu}\\
}

\maketitle

\begin{abstract}
Camera arrays (CamArrays) are widely used in commercial filming projects for achieving special visual effects such as bullet time effect, but are very expensive to set up.
We propose \emph{CamSwarm}, a low-cost and lightweight alternative to professional CamArrays for consumer applications. It allows the construction of a collaborative photography platform from multiple mobile devices anywhere and anytime, enabling new capturing and editing experiences that a single camera cannot provide.
Our system allows easy team formation; uses real-time visualization and feedback to guide camera positioning; provides a mechanism for synchronized capturing; and finally allows the user to efficiently browse and edit the captured imagery.
Our user study suggests that CamSwarm is easy to use; the provided real-time guidance is helpful; and the full system achieves high quality results promising for non-professional use.

\end{abstract}

\section{Introduction}
\label{sec:intro}

With carefully arranged spatial layouts and precise synchronization, camera array systems allow multiple cameras to work collaboratively to capture a moment in time of a scene from many view points.
It is the de facto solution for some popular visual effects such as the bullet time~\cite{BulletTimeBook}, or applications such as motion capture~\cite{MotionCaptureSurvey} and 3D scene reconstruction~\cite{CamArray3DReconstruct}.
It has been shown that an array of hundreds of low-cost cameras can produce high quality photographs that are even better than those taken by a professional DSLR~\cite{CamArray}\cite{Wilburn:2005}.
Given the pervasive popularity of consumer mobile cameras, it will be tremendously useful if we can extend the above ability to mass consumers.

\begin{figure}[t]
    \centering
    \includegraphics[width=0.49\textwidth]{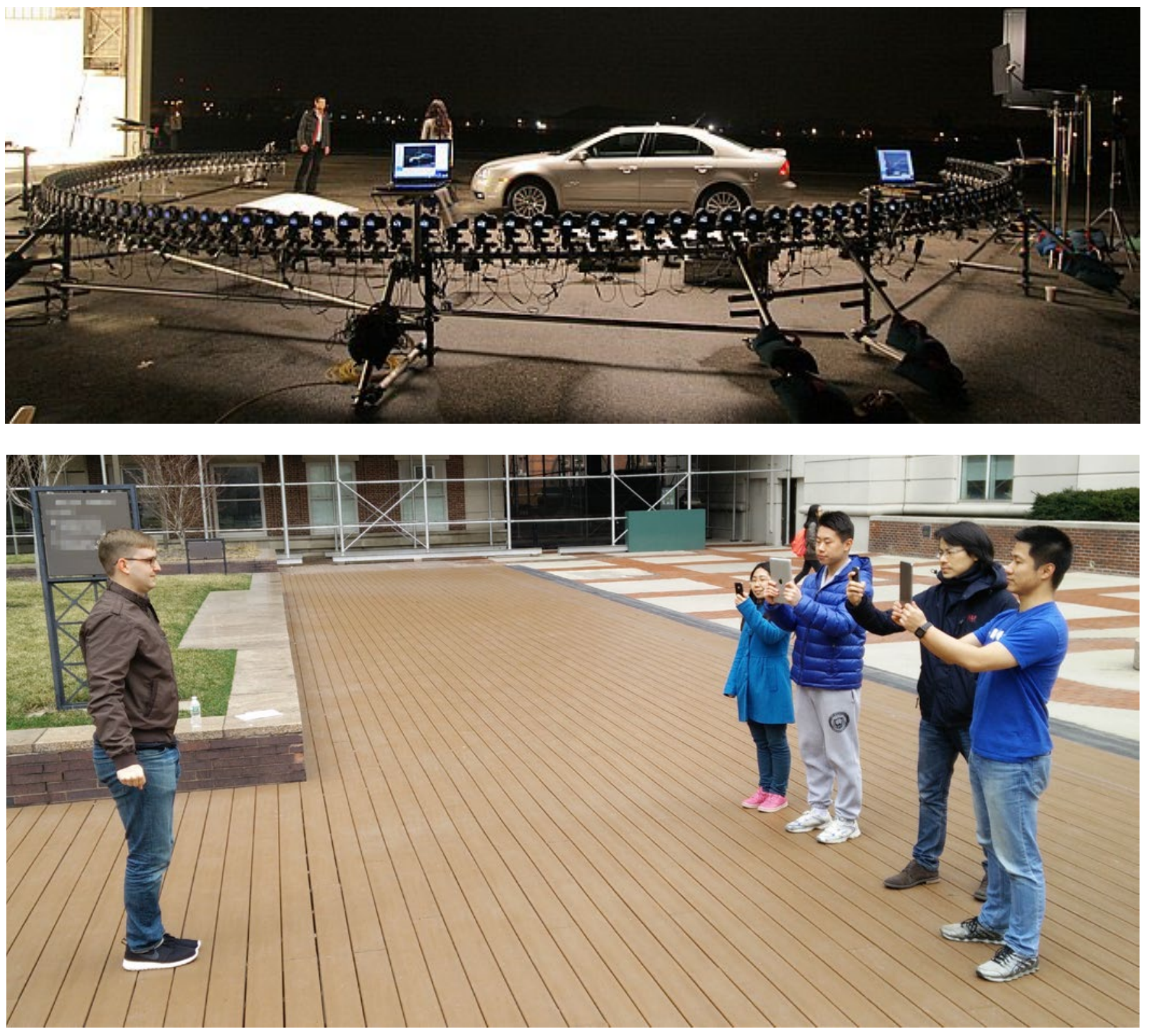}
    \caption{
        Top: A complex commercial camera array system (image credit: Breeze System).
        Bottom: We propose CamSwarm, a low-cost, smartphone-based instantaneous camera array that brings new collaborative imaging experience to consumers.
    }
    \label{fig:teaser}
\end{figure}

Unfortunately, existing CamArrays are far beyond the reach of ordinary consumers.
A professional array may include hundreds of DLSR cameras (see Figure~\ref{fig:teaser}), thus is very expensive to build.
Assembling a camera array involves placing each camera at carefully chosen locations and orientations, usually on giant rigs. 
Special trigger cables are used to wire the cameras to controlling computers for precise shutter synchronization.
Once set up, a camera array is hard to move without re-calibration. 
Recently there have been attempts to build camera arrays with consumer-grade devices such as GoPro cameras~\cite{GoProCamArray} or Nokia phones~\cite{NokiaCamArray}, however their total costs are still too high for casual enthusiasts, and the problem of complex and rigid setup remains the same. 

In this paper, we present a novel system, called ``CamSwarm'', and demonstrate novel imaging experience inspired by those provided by professional CamArrays. 
In contrast to traditional CamArrays needing hundreds of cameras and complex calibration,
our system allows nearby smartphones to quickly form a small-scale camera array, and synchronizes all devices to allow them to capture at the same time.
To help users position and direct their cameras, we provide a real-time interface to guide users to move, so that the cameras are properly spaced along a circle that centers at the target object, and are all pointed to it.
Using our system, an instantaneous, wireless smartphone camera array can be set up very quickly (e.g. within a minute).  

The key to the efficiency and agility of the proposed system is the collaborative interaction that keeps all users in the loop.
By showing the real-time rendered feedback to users, each participant can actively adjust the pose of his/her camera in order to help compensate any deficiency and improve the overall visual effect on the spot.
Such live feedback and control is not available in  conventional CamArrays, and is the fundamental feature that allows us to significantly reduce the complexity while retaining the CamArray-like imaging experience for general users. 
As applications, we demonstrate two ways to view and interact with the captured imagery:
(1) capture a highly-dynamic object from different view angles at the same time, and conveniently view all images on mobile devices using gyroscope data;
 and (2) capture multiple videos of an action from different angles, and render a bullet-time video.

To demonstrate the effectiveness of the system and the necessity of the provided features, we conduct a user study and compare it with simpler partial solutions by disabling features in the system.
Results show that our full system significantly outperforms these alternatives in terms of the perceptual quality of the final rendering results.   

Our research makes the following contributions:
\begin{itemize}
    \itemsep 0.1em
    \item the first system that forms an instantaneous camera array from multiple smartphones for consumer-level applications;
    \item a new real-time guidance interface to help users adjust camera configuration at capturing time in a live manner; 
    \item new interfaces for browsing and interacting with the captured multi-angle images or video.
\end{itemize}

\section{Related Work}
\label{sec:relatedwork}

\subsection{Multi-Camera Arrays}
Building multi-camera array systems using inexpensive sensors has been extensively explored.
These systems can be used for high performance imaging, such as capturing high speed, high resolution and high dynamic range videos~\cite{Wilburn:2004,Wilburn:2005}, or the light field~\cite{Venkataraman:2013}.
These systems however are not designed for multi-user collaboration or consumer applications. 
Some of them contain large-scale, fixed support structures that cannot be easily moved around, and they all require special controlling hardware for sensor synchronization.
Our approach, in contrast, uses existing hardware and software infrastructure (e.g. wireless communication) for camera synchronization, thus can be easily deployed to consumers. More importantly, user interaction plays a central role in our capturing process, but not in traditional camera arrays.    

\subsection{Collaborative Multi-User Interface}
Multi-user interface design is also a rising topic, especially in the HCI community.
The ``It's Mine, Don't Touch'' project~\cite{DontTouch} builds a large multi-touch display in the city center of Helsinki, Finland for the users to do massively parallel interaction, teamwork, and games. 
Exploration has also been done about how to enable multiple users to do painting and puzzle games on multiple mobile devices~\cite{MultiUserVideo}.
Our work is inspired by these works, but it also differs from these interfaces in that we make bullet time visual effect as the primary goal, providing users a means to obtain cool videos, which has its unique technical challenges and cannot be done with the interfaces mentioned above.

\subsection{Collaborations in Photography}

Collaboration has deep roots in the history of photography, as shown in a recent art project that reconsiders the story of photography from the perspective of collaboration~\cite{Collaboration}. This study created a gallery of roughly one hundred photography projects in history and showed how ``photographers co-labor with each other and with those they photograph''.   

Recently advances on Internet and mobile technologies allow photographers that are remote to each other in space and/or time to collaboratively work on the same photo journalism or visual art creation project.
For example, the 4am project~\cite{4am} gathers a collection of photos from around the world at the time of 4am, which has more than 7000 images from over 50 countries so far.
The ``Someone Once Told Me'' story~\cite{someone} collects images in which people hold up a card with a message that someone once told them.
In addition, shared photo album is an increasingly popular feature among photo sharing sites (e.g. Facebook) and mobile apps (e.g. Adobe GroupPix), which allows people participated in the same event to contribute to a shared album. 
All these collaborative applications focus on the sharing and storytelling part of the photography experience, but not the on-the-spot capturing experience.

PhotoCity is a game for reconstructing large scenes in 3D out of photos collected from a large number of users~\cite{Tuite:2011}.
It augments the capturing experience by showing the user the existing 3D models constructed from previous photos, as a visualization to guide the user to select the best viewpoint for taking a photo.
In this way, photos of different users can be combined together to form a 3D model.
However, this system is designed for {\em asynchronized} collaboration, meaning that photos are taken at different times.
In contrast, our system focuses on {\em synchronized} collaboration for applications that require all images/video/videoss to be taken at the same time. 

\subsection{Image based View Interpolation}

The video synthesize of bullet time effect heavily relies on the image-based view interpolation, which generates new views from the input images captured from other angles of the target.
One classical direction from $1990$s first reconstructs a 3D model of the target scene, and then does a 3D rendering from the new angle~\cite{ViewInterpolate98}\cite{shum2000review}.
There are other approaches which do not require explicit 3D reconstruction, such as lightfield rendering~\cite{LightFieldRendering} and plenoptic stitching~\cite{PlenopticStitching}, but usually require special equipments.
Considering the application scenario and the computational cost, we choose the Piecewise Planar Stereo~\cite{PhotoSynth}, which uses a hybrid method to achieve a balance in rendering quality and speed, as our backend.

\section{System Description}
\label{sec:approach}

\begin{figure*}[t]
    \centering
    \includegraphics[width=\textwidth]{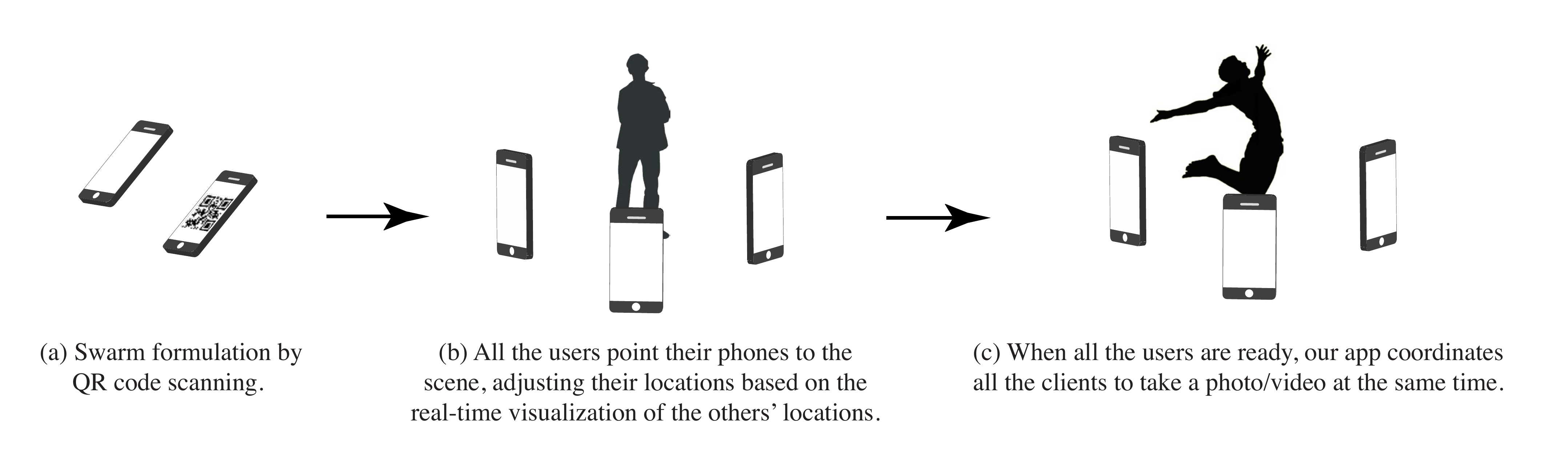}
    \caption{The workflow of forming a CamSwarm for capturing a scene.}
    \label{fig:workflow}
\end{figure*}

To assemble multiple cameras into a functional camera array, the following technical problems need to be addressed: (1) establish communication protocols among devices; (2) camera spacing and orientation adjustment; and (3) shutter synchronization for simultaneous capturing.

Shutter synchronization is obviously critical for capturing highly-dynamic objects. 
In traditional camera arrays, this is often achieved by wiring all cameras through trigger cables to a controlling computer.
Moreover, for achieving high quality view changing results in applications such as the bullet time effect, cameras need to be evenly spaced to have a balanced view angle coverage, and directed to point to the same focus spot.
This is achieved using specially designed camera array rigs in traditional systems.  

To make our system more ubiquitous, we resort to software solutions for addressing all problems, so that for end users, no additional hardware is required other than their smartphones. 
We adopt a server-client framework in a local WIFI network for cross-device communication, and develop a countdown protocol to allow a server program to trigger the shutter release event on all devices at the same time. 
For camera positioning, we provide a real-time interface to guide the users to move their cameras to achieve better spatial camera layout. 

A typical workflow of CamSwarm is show in \Fig{workflow}.
Next, we will describe each technical component in more detail. 

\subsection{Swarm Formation}

For cross-device communication, our system requires a local WiFi network.
A public WiFi or phone-hosted Ad-Hoc network is sufficient for the application.
The server program that coordinates all devices only needs to run on one of the participating smartphones, which we call the {\em host device} and is typically the first device that initiates the CamSwarm.

Specifically, a CamSwarm starts from a single user who first launches our mobile app. 
When it launches, a QR code that encodes the user's IP address is created and shown on the screen.
Other users can scan this QR code to join the CamSwarm, by talking to the server program at the obtained IP address. 
Once a swarm is formed, the same QR code will display on all participating devices, allowing more users to scan and join.
This QR code propagation strategy allows multiple users to quickly form a team. 

\subsection{Collaborative Camera Positioning}

\subsubsection{Guidelines for Camera Positioning}

In many camera array applications, one needs to smoothly interpolate between cameras at adjacent view angles to create a steady panning effect centered on the target scene/object.
The quality of the final output largely depends on the spatial distribution of the participating cameras. 

Based on the general principles of view interpolation algorithms, we propose the following guidelines for camera positioning to help improve the quality of the final effect: 
\begin{itemize}
    \itemsep 0.1em
    \item adjacent views should have sufficient scene overlap, allowing multi-view image matching to work reliably;
    \item the angle differences between adjacent cameras should be roughly the same to ensure smooth view interpolation, because large view angle difference is likely to cause visual matching failures;
    \item the main target should have similar sizes and relative positions in images captured by different cameras.
\end{itemize}

Although these guidelines seem to be straightforward, as we will show in the user study later, without proper visual guidance, it is hard for a group of users to achieve good spatial camera layout.
One major problem is that the angle of the {\em view change} between two users is hard to be directly measured by human perception.
For close-up shots, slightly moving away from the neighboring user may cause a large enough view change to break image matching.
On the contrary, for far-away objects, a seemingly enough movement may actually be insufficient to maintain the desired angle difference.
Furthermore, if different types of devices are used in the same array (i.e. some iPhones with some iPads), the field of view of different users may be different, requiring proper adjustment of the distances to the target object accordingly.
We thus design a real-time interface to guide the users to achieve a good spatial layout quickly.   

\begin{figure}[t]
    \centering
    \includegraphics[height=9cm]{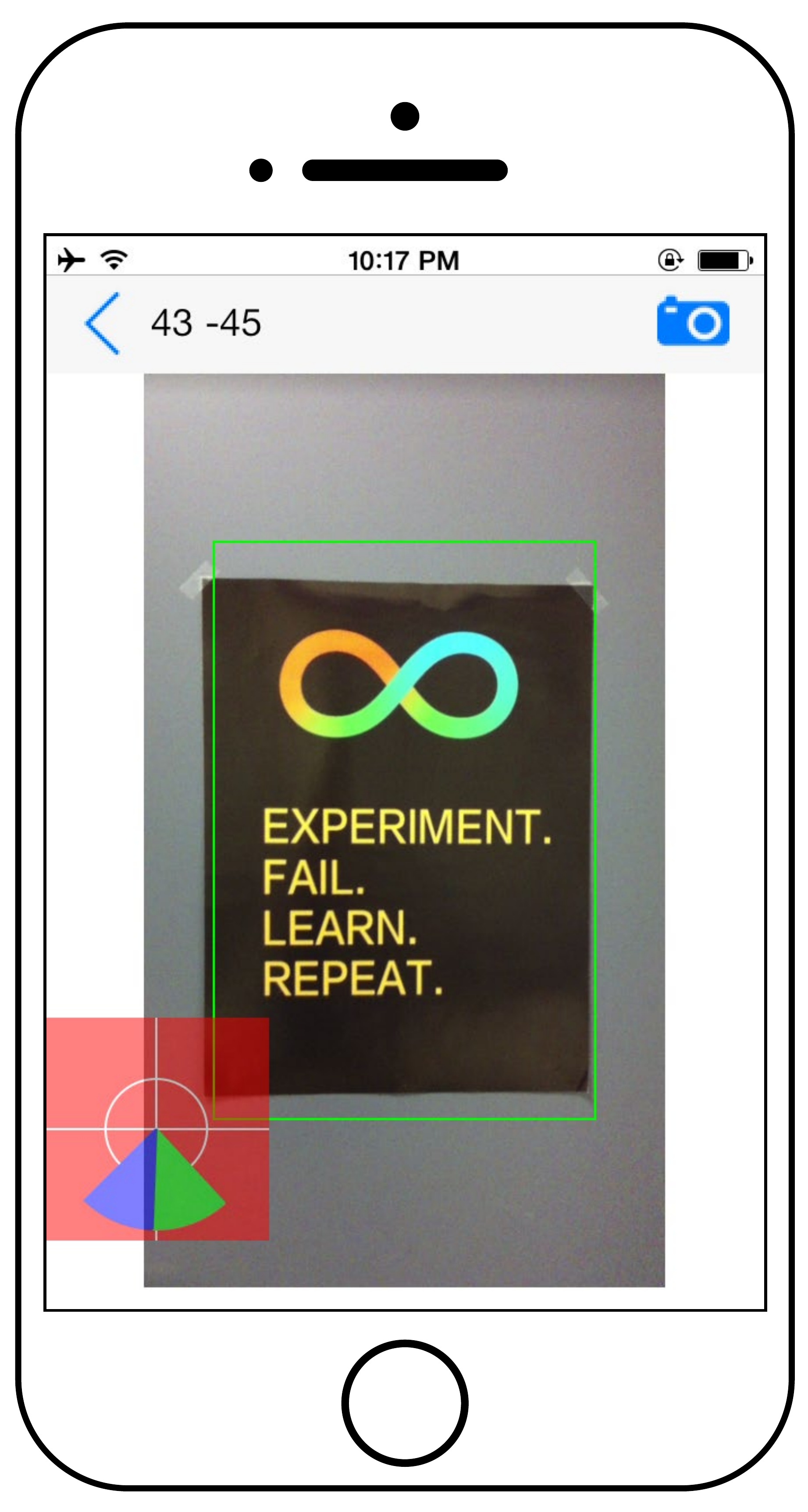}
    \caption{
        The capturing interface. The green guiding box is specified by the host for adjusting the size and position of the object. The compass in the bottom-left corner shows the angular distribution of all the cameras.
        The numbers in the top-left corner show the angle values between neighboring cameras for fine position tuning.
    }
    \label{fig:compass}
\end{figure}

\subsubsection{Interface for Camera Positioning} 
A screenshot of our real-time interface is shown in \Fig{compass}. The main window shows the viewfinder image with a guiding box overlaid on the top (shown in green). The size and position of the guiding box is specified by one of the users in the array, so that it tightly encloses the target object on his/her screen. Once specified, the guiding box is then transmitted to all other devices, and other users can adjust their own  camera orientations and the camera-object distances to fit the main target into the guiding box. 
We have also experimented with computer-vision-based methods to automatically compute the object size variations across devices, but found them to be fragile for objects that are lack of salient textures. 

On the lower left corner of the interface, we present a compass-like visualization of the angular distribution of all devices, so that the users can intuitively determine how to move to achieve an even angular distribution. The visualization is self-centric on each device: the target object is always at the center of the compass and the current device is always at the south end. This is implemented using both gyroscope and compass readings of the devices. 
Although it is not possible to directly determine the users' locations relative to the filming target from sensor data, if the users all point their cameras to the target object,  the gyroscope readings can reflect the users' relative angles. 
While gyroscopes can only provide relative bearing from a reference, digital compass can provide a reliable reading of magnetic north after proper calibration, which enables us to compare the gyroscope readings across devices.

Specifically, each device first reads from the gyroscope and the compass, and then sends the relative bearing between the yaw of the device and the magnetic north to the server program.
The server program then computes the relative yaw of each device against the host device, and broadcasts this information.
It is worth noting that a right-facing device is actually in the left part of the array, assuming that all cameras point to the same target object.
To properly visualize the relative camera locations, the server sends the negative of the relative yaw to each device.

\subsection{Synchronized Capture}
Once the cameras are properly distributed and adjusted, all shutters should be triggered at the same time for synchronized capture.
An straightforward solution is to let the host device broadcast a signal when the host user taps the ``capture'' button.
However, after extensive testing we found such a solution has two problems.
Firstly, the signal packet may get lost during transmission, result in a failed capture action.
Secondly, different devices usually have different latency in the network, thus they receive the signal at slightly different times.
These problems are especially severe in the wild, such as public WiFis or mobile-hosted Ad-Hoc networks.
In our experiments, we found that the signal lost probability can go up to 50\%, and the signal delay can extend to more than a second.

To avoid these problems, we use a postponed capture scheme. 
The host device is set to keep broadcasting countdown signals from $5$ seconds before the scheduled capture.
Each signal carries the remaining waiting time in milliseconds.
Therefore for each other device, the capture action will eventually be triggered as long as one of the many broadcasted packets is received, and the actual latency is the minimum transmission latency among all the received packets. In our experiments, we found that after adopting this scheme, missed capture rarely happens, and the average latency is reduced to roughly $50$ms. When the actual capturing starts, each device can capture either a single image, or a short video sequence of a fixed length, depending on the instruction embedded in the countdown signal from the host device. 

\begin{figure}[t]
  \subfigure[]{
    \includegraphics[width=0.21\textwidth, trim=0 0 5 0]{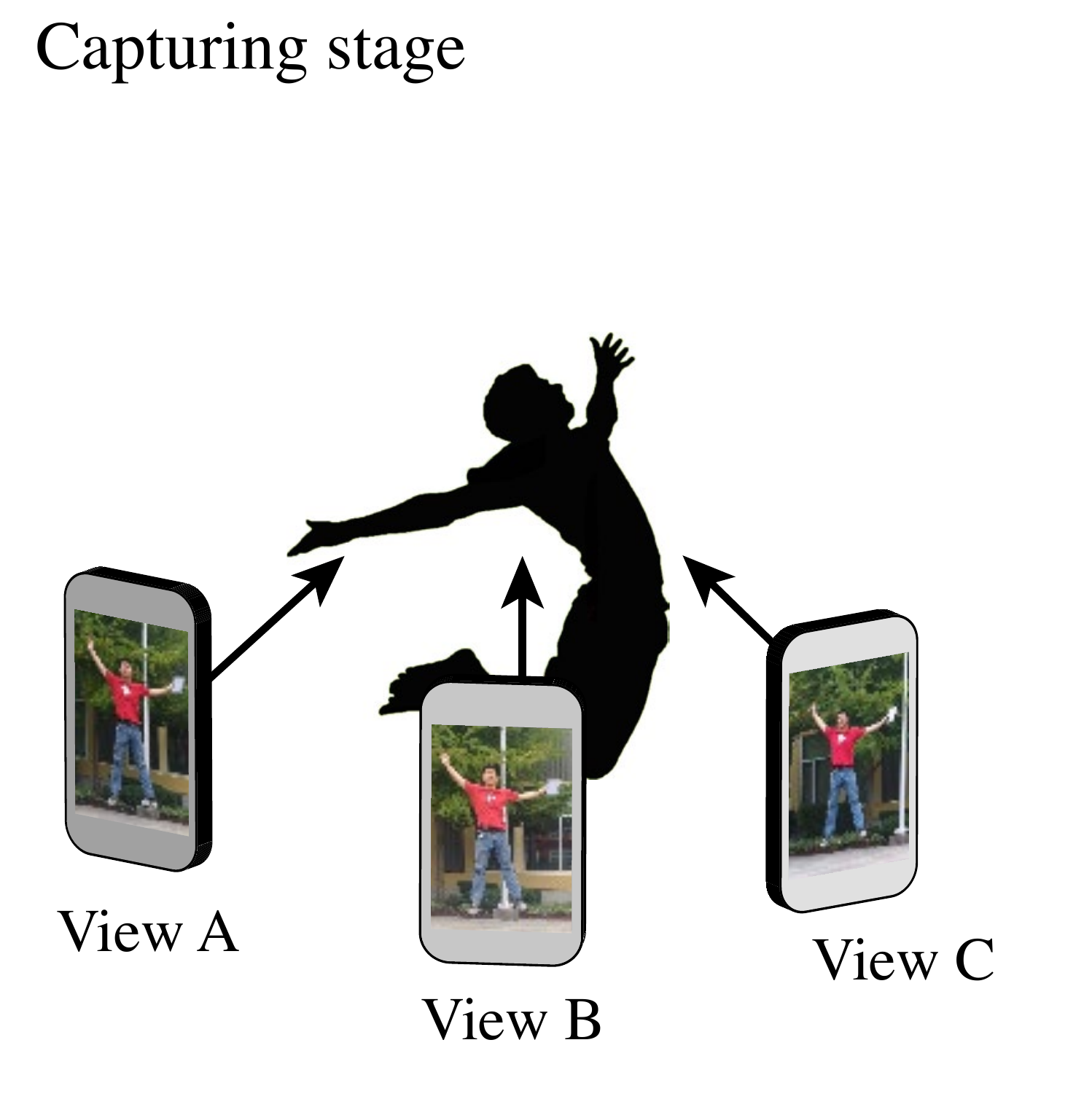}
    \label{fig:render_photo:scene}
  }
  \hfill
  \subfigure[]{
    \includegraphics[width=0.23\textwidth, trim=0 0 5 0]{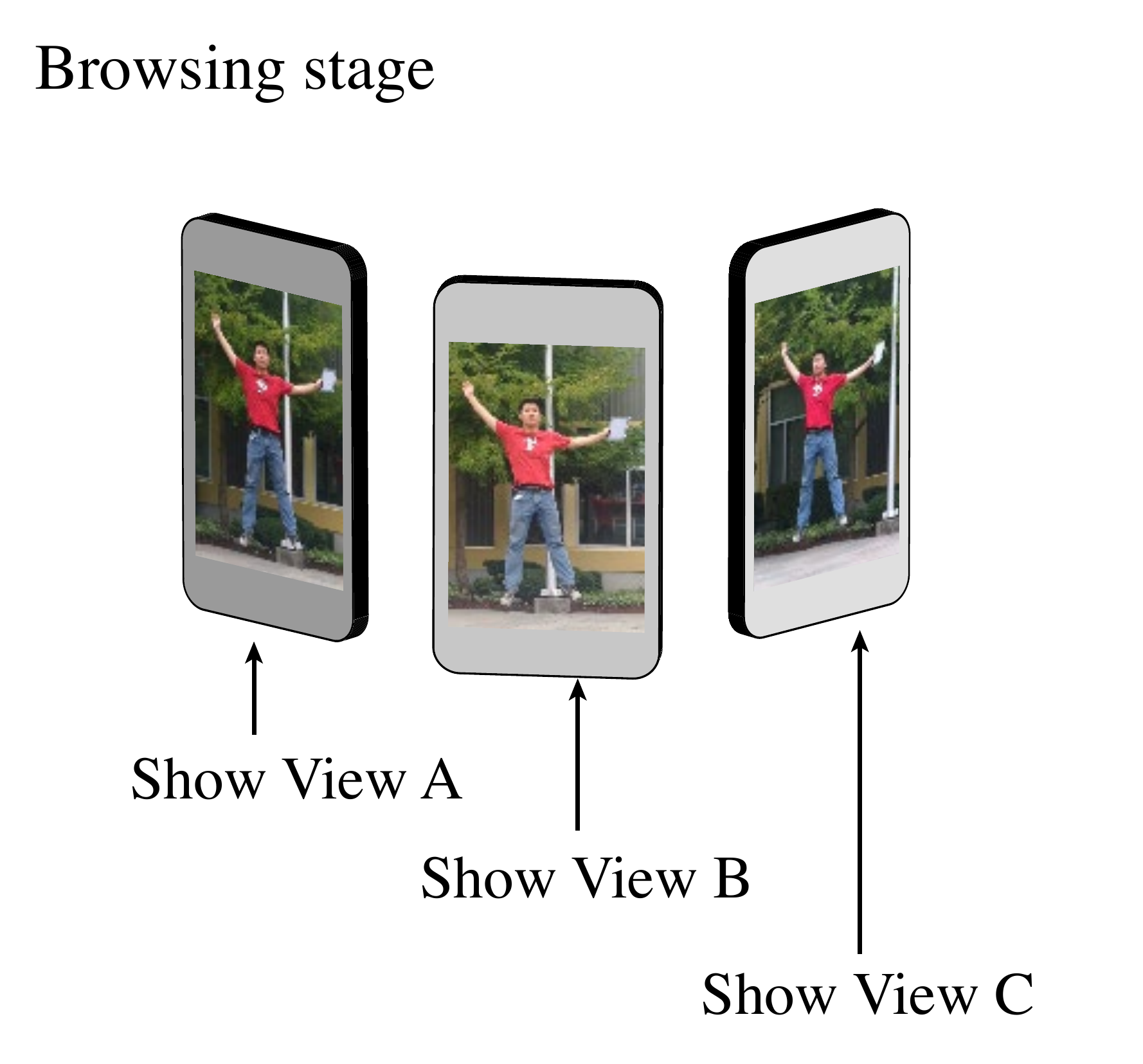}
    \label{fig:render_photo:phone}
  }
  \caption{Gyroscope-based browsing interface. (a) Three devices A, B and C are used to capture a scene. (b) At the browsing stage, each user can tilt his/her device to switch to another view.}
  \label{fig:render_photo}
\end{figure}

\section{Viewing and Editing Interfaces}
\label{sec:render}

A CamSwarm produces a set of images or video sequences that are captured from different view points at the same time. One can simply present all the raw imagery to each user for consumption, but this naive viewing experience does not take advantage of the spatial layout of the camera array. In our system we present two new interfaces for browsing and editing the captured camera array data.

\begin{figure*}
  \subfigure[An overview of the interface.]{
    \includegraphics[width=0.32\textwidth, trim=0 0 5 0]{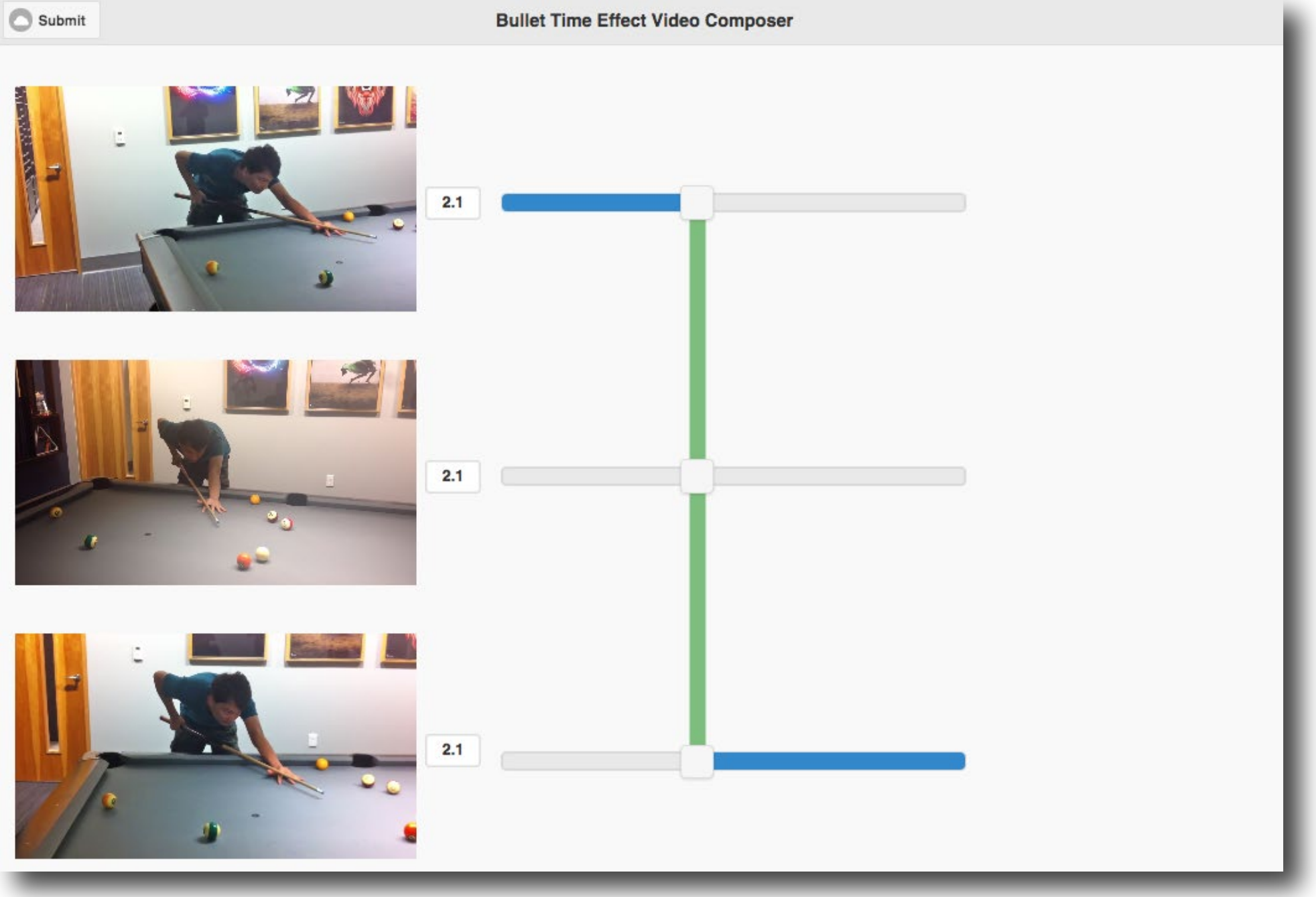}
    \label{fig:render_video:interface}
  }
  \hfill
  \subfigure[The user can select a certain view and use the interlocked cursor to seek for a proper view transition point.]{
    \includegraphics[width=0.32\textwidth, trim=0 0 5 0]{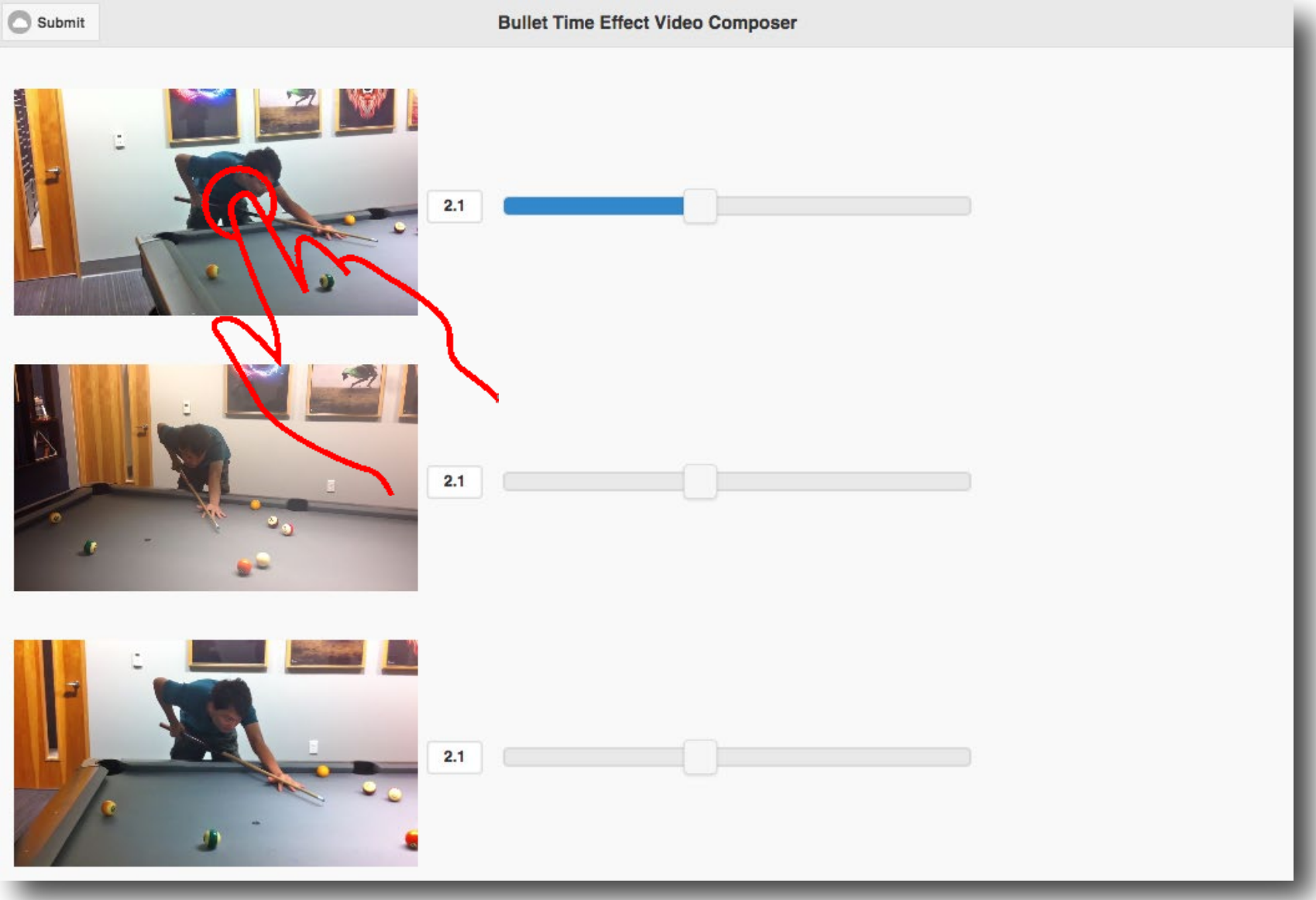}
    \label{fig:render_video:cursor}
  }
  \hfill
  \subfigure[The user can tap the video preview of another view to trigger a view transition.]{
    \includegraphics[width=0.32\textwidth, trim=0 0 5 0]{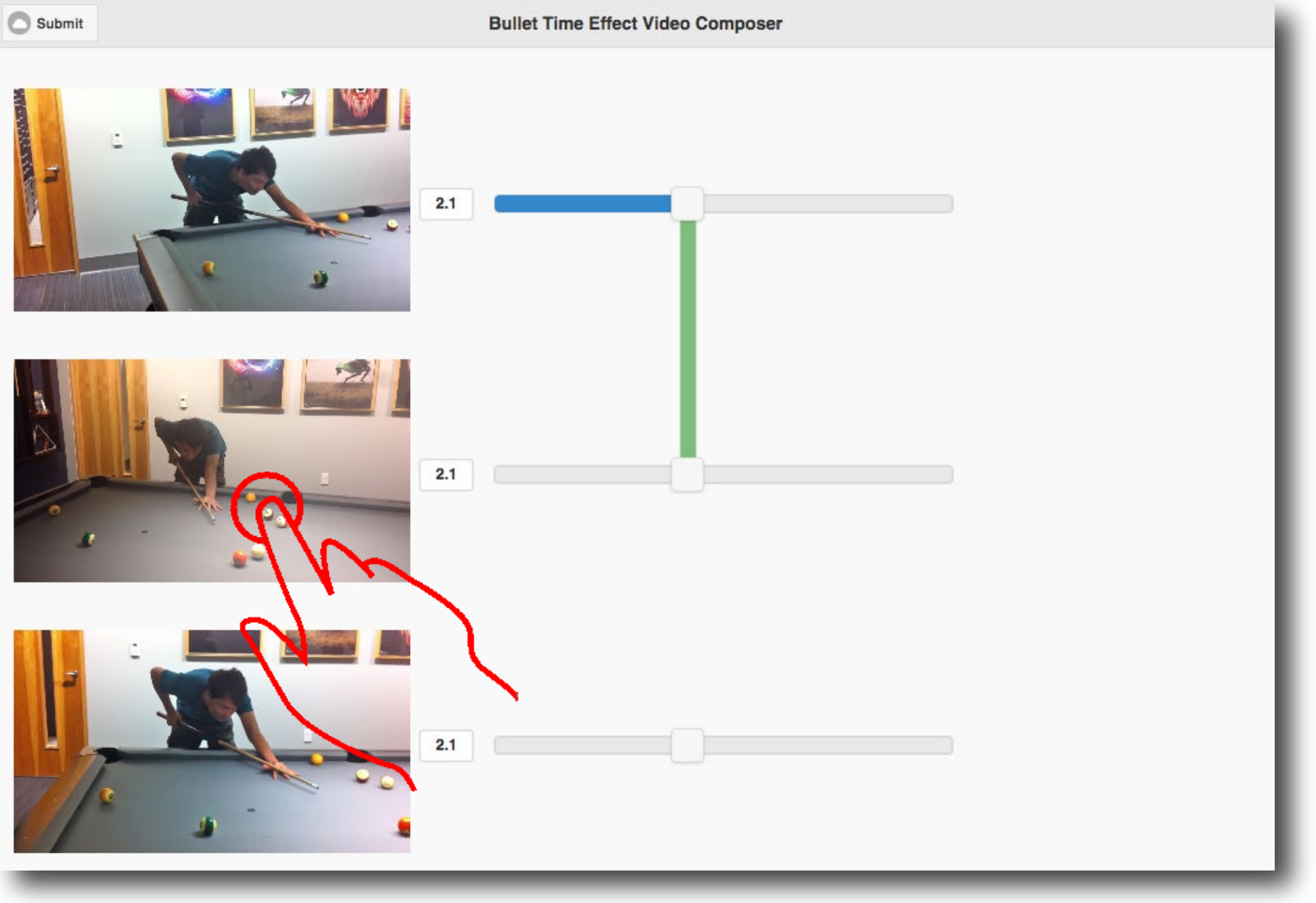}
    \label{fig:render_video:transition}
  }
  \caption{Our interface for creating a bullet time video.}
  \label{fig:render_video}
\end{figure*}

\subsection{Gyroscope-based Browsing}

We first provide a natural way to browse the captured scene images on mobile devices by tilting the phone.
As shown in \Fig{render_photo}, once captured, the user at View B can not only see the image captured from this view point, but also switch to see the images captured from View A and C, by tilting the phone to change the view angle in either direction.

To implement this interface, we establish a mapping from the gyroscope reading to which view to show in the browsing stage.
Because the relative yaws of each phone are known at capturing stage, we first normalize the angles to center them at 0, and then build a Voronoi graph from the relative yaws.
In the browsing stage, when the user tilts the phone, we compute the relative yaw from the initial orientation when the browsing begins, and then fit it into the Voronoi graph to obtain the view with the closest relative yaw.

\subsection{Bullet-Time Video Composition}

The key advantage of using a camera array is that we can create a bullet time video from all the captured video sequences. In our system we provide an easy interface to guide the user to create such videos. 
Compared with professional arrays with hundreds of devices, CamSwarms typically have far fewer cameras, therefore it is necessary to use view-based rendering techniques to generate smooth transitions among different views, an extensively studied problem in computer vision. 
In our system we use the piecewise planar-stereo-based rendering~\cite{PhotoSynth} for view interpolation, given its fast speed and relatively good results. 

Our HTML5 interface for creating a bullet time video is shown in~\Fig{render_video:interface}. 
It consists of two panels: a video preview panel on the left, showing all captured sequences as a list; and a timeline editing panel on the right, showing the transitions from one view to another that are specified by the user. The user first selects a video to begin with, and then selects the transition point by dragging the corresponding timeline. The selected portion of the video that will be included in the final composition is shown as a thick blue line in~\Fig{render_video:cursor}. 
Note that since the captured video sequences are synchronized, a single timeline controls all preview videos. 
At this point, the user can select another view by tapping, indicating an insertion of a  view transition, as shown as a vertical green line in~\Fig{render_video:transition}. This process continues until the user is satisfied or reaches the end of the timeline. The final video is then rendered on a server and sent back to the user momentarily.

\section{Evaluation}
\label{sec:exp}

We implement the proposed system as an iPhone app.
A video demonstrating the system in action and the final synthesized videos is available at \url{https://www.youtube.com/watch?v=LgkHcvcyTTM}.
For system evaluation, we have conducted two user studies to answer two main questions: (1) in the capturing stage, whether the proposed visual guidance can help users achieve better camera positioning; and (2) if the system can generate higher quality results than simpler alternatives.


\subsection{Study I: Evaluation on Real-Time Visual Guidance}

We first explore whether the propose visual guidance shown in~\Fig{compass} can help camera positioning through a user study. 

\subsubsection{Evaluation Settings}
We set up an indoor experimental scene with a single foreground object, and invite four subjects to take picture of it in every session.
The subjects are asked to point their phones to the foreground object, and comply with the guidelines mentioned in Section System Description, i.e. to scatter evenly around the object, and to ensure the object have similar size in the photo/viewfinder.
Each session consists of two trials, one without the real-time guidance, and one with.
The subjects are asked to leave and re-enter the experiment site and switch orders after the first trial, so their positions from the previous trial will not affect the second one. For the trail with the visual guidance, we briefly explain the two visual components, the green guiding box and the angular compass beforehand.  

We choose to use a planar object as the foreground in this experimental scene, so that the distances and the relative angles to the object can be reliably computed using computer vision techniques from the captured photos (note that in real applications our system does not have any shape assumption on the object).
This allows us to compute the standard derivation in the distances and relative angles to the foreground object as quantitative evaluation protocols.
To prevent the factor of system familiarity affecting the result, in half of the sessions the real-time guidance interface is used first, while in the other half the order is reversed.
The subjects are allowed and encouraged to use both verbal and gesture communication in the entire process, regardless of which system they are using.

After collecting all the photos, we manually label the corners of the planar foreground object in each photo, and use the pinhole camera model to compute the angles between the normal direction of the object and the principal direction of each camera.
And then the angles between adjacent user pairs are computed.
The final evaluation protocol is the relative standard derivation of the angles, which is defined as the standard derivation divided by the average of the angles. This is to avoid the preference to  smaller angles.
The relative standard deviation is also computed on the extracted foreground object size (in pixel).

\subsubsection{Result and Discussion}
\begin{table}
    \centering
    \begin{tabular}{c|c|c}
        \hline
        & Size & Angle \\
        \hline 
        With preview & $0.05$ & $0.097$ \\
        Without preview & $0.034$ & $0.272$ \\
        \hline
    \end{tabular}
    \caption{Quantitative results of the user study on the real-time visual guidance for camera positioning.}
    \label{tbl:preview}
\end{table}
We invite $20$ subjects in 5 groups to participate in this study.
The computed relative standard derivation is averaged across different groups, and are reported in \Table{preview}.
The results show that with the help of the real-time visual guidance, the relative deviation of the angles are reduced significantly to about one third.
However, the designed interface does not help in the distance measurement, the relative deviations are both small with and without the guiding box.
We think this is largely due to the simple scene setup in this experiment.
Given that there is only a single foreground object, we have observed that in the trails without the green guiding box, subjects tend to use the border of the screen as an implicit guidance, so that the object roughly occupies the whole screen. 
We believe such a feature is more useful in more complicated real-world scenes, where the users may only want to keep the object at a specific location in the final images for a better scene composition. 

Overall, considering that an even angular camera distribution is critical for high quality view interpolation, and the size difference can be more easily compensated by resizing in post-processing, we conclude that the proposed interface can help users achieve better camera positioning. 

\subsection{Study II: Evaluation on Visual Quality}

To evaluate whether better camera positioning and synchronized capturing can lead to higher visual quality output, 
we conduct subjective evaluation on the output of our system against that of several baseline approaches.

\subsubsection{Baselines}
We compare our full approach with the following baseline methods:
\begin{itemize}[noitemsep]
    \item \emph{No Sync, No Guidance.} We disable both synchronized shutter and visual guidance for camera positioning in our system to create this baseline. It is equivalent to capturing the scene with regular camera apps, and using the proposed backend to generate the final visualization.
    \item \emph{Sync, No Guidance.} The proposed system with synchronized shutters, but without real-time guidance.
\end{itemize}

\subsubsection{Testing scenes}

For this study, we choose three realistic scenes as the representative use cases of the CamSwarm system:
\begin{itemize}[noitemsep]
    \item \emph{Billiard.} An indoor scene of one person playing billiard in a room.
    \item \emph{Jump.} A dynamic scene of a person jumping in front of a camera.
    \item \emph{Martial Arts.} A highly dynamic scene of a person performing martial arts in front of camera.
\end{itemize}
These baselines are implemented as special modes in our system. For each scene, we recruit one actor to perform an action, and one group of subjects in Study I to take videos using each of the comparison method, in random orders.
Before capturing, a brief tutorial session is provided about how to use the app. 

\subsubsection{Questionnaire and Subjects}

After collecting all data, the same settings and processing procedures are performed to generate final visual results in the two presentation forms: gyroscope-based browsing and bullet time video. 
These results are presented to $30$ different subjects for visual quality evaluation. 

We first ask the subjects to evaluate the bullet time video results, For each scene, the three output videos generated by three different systems are played side-by-side in a random order, and a questionnaire is provided to evaluate different aspects of the video.
For each question, we ask the subjects to rate from $1$ to $5$, with $1$ as totally disagree and $5$ as totally agree.
\begin{enumerate}[noitemsep]
    \item I feel the main object stays stable in the video (e.g. it barely moves).
    \item I feel the artifacts are acceptable.
    \item I feel the time is frozen.
    \item I feel the transition is natural, as if I am walking around the object viewing it.
    \item Overall, how do you like the result? ($1$ to $5$)
    \item If you can download an app on your phone to create videos like this with some friends, would you do it? ($1$ is least likely, and $5$ is most likely)
\end{enumerate}

For gyroscope-based browsing, the subjects are required to use the interface on a smartphone for a fully fledged experience. because this form of presentation does not have view transitions, only Question 5 is asked.  

We also ask the subjects participating in the capturing sessions a few more questions after they have finished all the tasks, with the same scoring system about agreeness:
\begin{itemize}[noitemsep]
    \item The user interface is easy to learn.
    \item The user interface is easy to use.
\end{itemize}

\subsection{Results and Discussion}

We now discuss the collected scores and feedback from the subjects.

{\em Robustness. }First of all, not all of the footage captured by the subjects can be successfully stitched to create bullet time video by the computer vision backend. 
In order to obtain a successful bullet time video, the subjects tried $3.67$ times in average using the \emph{Sync, No Guidance} system, and tried $4$ times in average using the \emph{No Sync, No Guidance} system.
In contrast, all the user groups are able to produce a view-able bullet time video in the first trial, with $1.33$ trials to achieve the first successful capture in average (including the final successful trial).
This shows that the proposed system has a much higher successful rate at capturing and creating a bullet time video than alternative approaches.

\begin{figure*}[t]
    \centering
    \includegraphics[width=0.9\textwidth]{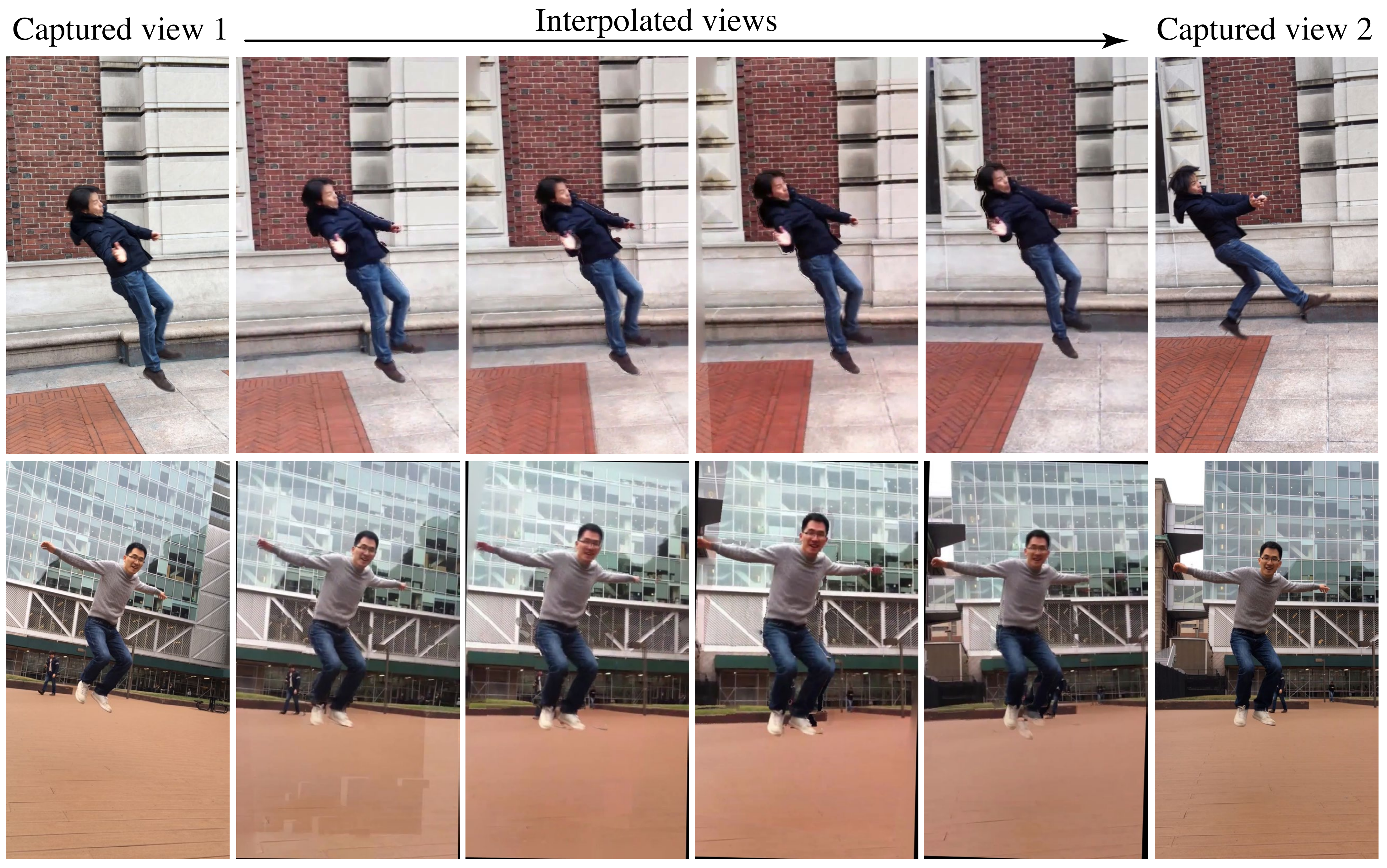}
    \caption{
        Example bullet time videos generated by our system. 
        For each scene, the left and right columns show two frames captured by two adjacent cameras, and all frames in-between are synthesized by the employed view interpolation method.
    }
    \label{fig:examples}
    \vspace{-0.5cm}
\end{figure*}

\begin{figure}[t]
    \centering
    \includegraphics[width=0.45\textwidth]{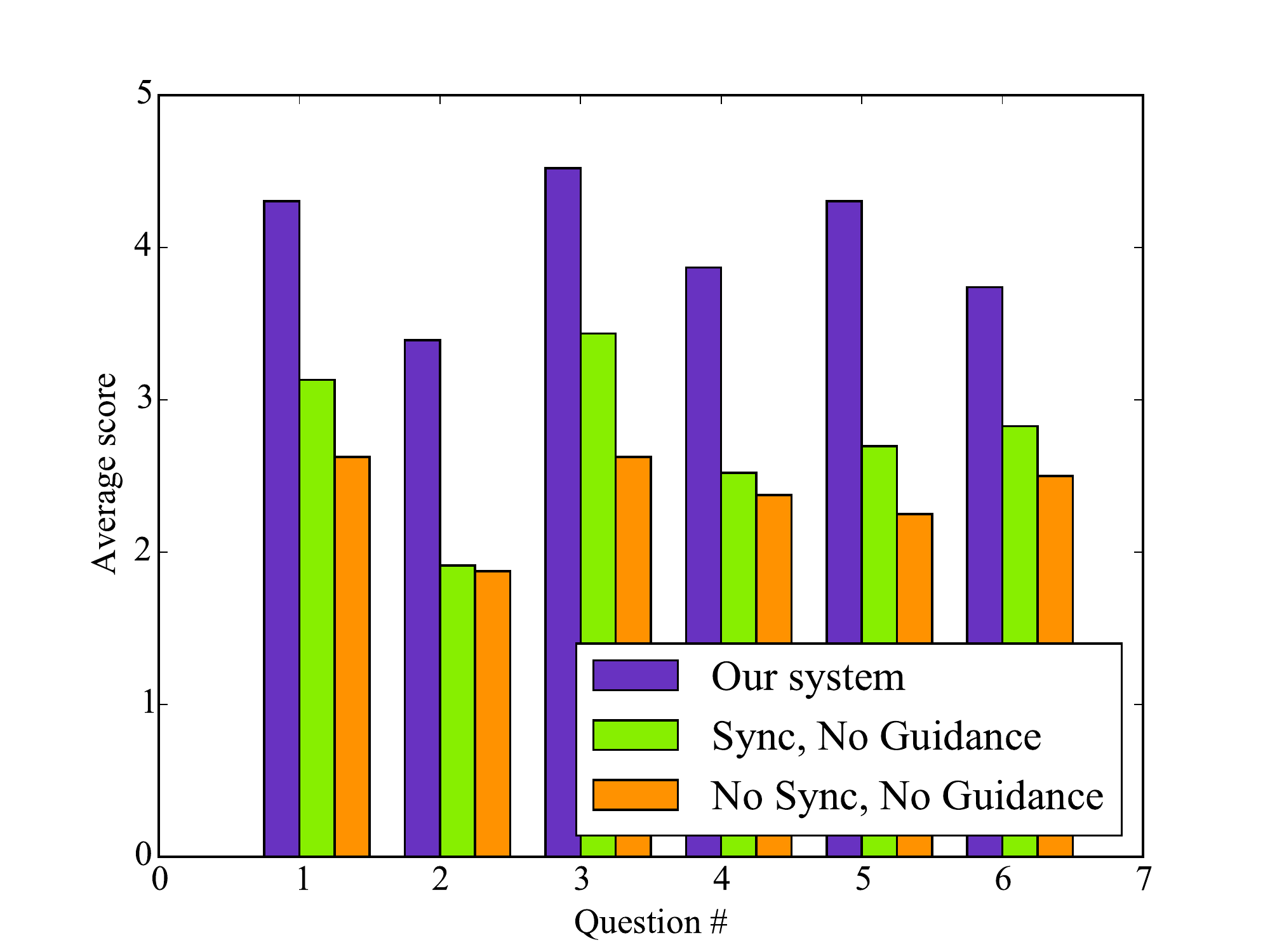}
    \caption{Results of the user study questionnaire on bullet time video. }
    \label{fig:userstudy}
    \vspace{-0.5cm}
\end{figure}

{\em Performance. }We collect the user scores of the bullet time videos, and plot the average scores in \Fig{userstudy}.
The suggest that the viewers are more satisfied with the results generated from the proposed system, consistently in all aspects.
The real-time visual guidance contributes to a significant portion of the visual quality improvement, further supporting our claim that good camera positioning is critical for achieving more visually appealing results. 
Synchronous shutter also contributes to improve the stability (Question $1$) and the feeling of the time is frozen (Question $3$).
Most subjects also show interest to use the system in practice, indicated by the relatively high score of Question 6. 
This is encouraging, as it suggests that even though our results contain visual artifacts, they may already be enjoyable to casual users in some cases.  

{\em Artifacts. }Compared with other aspects, the subjects are less satisfied with the view interpolation artifacts in the result video (Question $2$).
Example frames of the generated videos are shown in \Fig{examples}, where the interpolated images in the bottom row contains more artifacts than those in the top row.
It suggests that the specific view interpolation method we have chosen performs well when the scene mainly consists of planar surfaces, and have more difficulties in segmenting and matching scene regions when the scene is too cluttered.
Given that view interpolation is an active research area in computer vision, we expect that newer and better methods will soon emerge, which can be easily plugged into the system to significantly reduce artifacts. 

{\em Other results. }The gyroscope-based browsing received an average rating of $3.4$ out of $5$ using the full system, higher than the ratings of the baseline approaches {\em Sync, No Guidance} and {\em No Sync, No Guidance}, which are $3.1$ and $2.9$, respectively. The results suggest that even in this straightforward form of presentation, better camera positioning and synchronization can lead to better browsing experiences. 
For the ease of learn and use questions, the subjects participating in the capture stage give an average rating of $4.4$ and $4.2$ respectively.


{\em Group size. }In our experiments, we have found that a group of four users usually requires less than one minute to set up a swarm and capture a scene, after some initial training and practice.
The QR-code based pairing and real-time visual guidance play important roles on achieving quick setup and capturing.

We have also tested the reliability of the system when more users participate in a swarm.
No obvious performance degeneration is observed when eight devices work together.
Theoretically it is possible for the system to add more users, but we found that a group size larger than eight is often not helpful in practice, because some of the users may inevitably appear in some other users' camera views when more users present at the scene.

\section{Conclusion}
\label{sec:conclusion}

We have presented a collaborative mobile photography platform called CamSwarm, as a low-cost, consumer-level substitute to  professional camera arrays. 
CamSwarm allows smartphone users to dynamically form a collaborative team within a minute using a QR code propagation mechanism. 
It provides real-time location/orientation visualization to guide the users to better positioning their cameras in the capturing process, and presents intuitive interfaces for browsing the captured images and creating bullet-time video. 
Preliminary user study results suggest that the system can help users achieve higher quality output than simpler alternatives, and is easy and fun to use. 

As future work, we plan to investigate better view interpolation methods to reduce the visual artifacts in the bullet time video. We also plan to explore new applications of the collaborative photography platform, such as drone-based bullet time videos and panoramic image and video stitching.

\balance

\bibliographystyle{acm-sigchi}
\bibliography{main}
\end{document}